\documentclass[12pt]{article}
\usepackage{graphicx}

\newcommand\pubnumber{based on LHCb-CONF-2014-004}
\newcommand\pubdate{\today}

\def\napoli{on behalf of the LHCb collaboration}
\def\support{\footnote{CERN, Geneva, Switzerland}}

\def\Title#1{\begin{center} {\Large #1 } \end{center}}
\def\Author#1{\begin{center}{ \sc #1} \end{center}}
\def\Address#1{\begin{center}{ \it #1} \end{center}}

\newcommand\pubblock{\rightline{\begin{tabular}{l} \pubnumber\\
         \pubdate  \end{tabular}}}

\newenvironment{Presented}{\begin{quotation} \begin{center} 
             PRESENTED AT\end{center}\bigskip 
      \begin{center}\begin{large}}{\end{large}\end{center} \end{quotation}}

\usepackage{microtype}
\usepackage{lineno}  
\newcommand*\patchAmsMathEnvironmentForLineno[1]{%
\expandafter\let\csname old#1\expandafter\endcsname\csname #1\endcsname
\expandafter\let\csname oldend#1\expandafter\endcsname\csname
end#1\endcsname
 \renewenvironment{#1}%
   {\linenomath\csname old#1\endcsname}%
   {\csname oldend#1\endcsname\endlinenomath}%
}
\newcommand*\patchBothAmsMathEnvironmentsForLineno[1]{%
  \patchAmsMathEnvironmentForLineno{#1}%
  \patchAmsMathEnvironmentForLineno{#1*}%
}
\AtBeginDocument{%
\patchBothAmsMathEnvironmentsForLineno{equation}%
\patchBothAmsMathEnvironmentsForLineno{align}%
\patchBothAmsMathEnvironmentsForLineno{flalign}%
\patchBothAmsMathEnvironmentsForLineno{alignat}%
\patchBothAmsMathEnvironmentsForLineno{gather}%
\patchBothAmsMathEnvironmentsForLineno{multline}%
}
\usepackage{xspace} 
\usepackage{amsmath} 
\usepackage{amssymb}
\usepackage{amsfonts}
\usepackage{upgreek} 
\usepackage{graphicx}  
\usepackage{colortbl}
\graphicspath{{./figs/}} 
\usepackage[abs]{overpic}
\usepackage{hyperref}    
\usepackage[all]{hypcap} 
\usepackage{cite} 
\usepackage{mciteplus}
\usepackage{array}
\usepackage{ragged2e}
\newcolumntype{P}[1]{>{\RaggedRight\hspace{0pt}}p{#1}}
\usepackage{ifthen} 
\newboolean{pdflatex}
\setboolean{pdflatex}{true} 
\newboolean{articletitles}
\setboolean{articletitles}{true} 
\newboolean{uprightparticles}
\setboolean{uprightparticles}{false} 

\def\lhcb {LHCb\xspace}
\def\ux85 {UX85\xspace}

\ifthenelse{\boolean{uprightparticles}}%
{

 \def\Ppi         {\ensuremath{\uppi}\xspace}

 \def\PDelta      {\ensuremath{\Delta}\xspace}                 
 \def\PXi      {\ensuremath{\Xi}\xspace}                 
 \def\PLambda      {\ensuremath{\Lambda}\xspace}                 
 \def\PSigma      {\ensuremath{\Sigma}\xspace}                 
 \def\POmega      {\ensuremath{\Omega}\xspace}                 
 \def\PUpsilon      {\ensuremath{\Upsilon}\xspace}

 \def\PB      {\ensuremath{\mathrm{B}}\xspace}                 
                  
 \def\PD      {\ensuremath{\mathrm{D}}\xspace}

 \def\PK      {\ensuremath{\mathrm{K}}\xspace}

 \def\Pi      {\ensuremath{\mathrm{i}}\xspace}

 \def\Ps      {\ensuremath{\mathrm{s}}\xspace}

}
{

 \def\Ppi         {\ensuremath{\pi}\xspace}

 \mathchardef\PDelta="7101
 \mathchardef\PXi="7104
 \mathchardef\PLambda="7103
 \mathchardef\PSigma="7106
 \mathchardef\POmega="710A
 \mathchardef\PUpsilon="7107
                  
 \def\PB      {\ensuremath{B}\xspace}                 
                  
 \def\PD      {\ensuremath{D}\xspace}

 \def\PK      {\ensuremath{K}\xspace}

 \def\Pi      {\ensuremath{i}\xspace}

 \def\Ps      {\ensuremath{s}\xspace}

}

\def\g      {\ensuremath{\Pgamma}\xspace}

\def\squark    {\ensuremath{\Ps}\xspace}

\def\pion  {\ensuremath{\Ppi}\xspace}

\def\pip   {\ensuremath{\pion^+}\xspace}

\def\kaon  {\ensuremath{\PK}\xspace}
  \def\Kbar  {\kern 0.2em\overline{\kern -0.2em \PK}{}\xspace}

\def\Kz    {\ensuremath{\kaon^0}\xspace}
\def\Kzb   {\ensuremath{\Kbar^0}\xspace}
\def\KzKzb {\ensuremath{\Kz \kern -0.16em \Kzb}\xspace}
\def\Kp    {\ensuremath{\kaon^+}\xspace}
\def\Km    {\ensuremath{\kaon^-}\xspace}

\def\KpKm  {\ensuremath{\Kp \kern -0.16em \Km}\xspace}
\def\KS    {\ensuremath{\kaon^0_{\rm\scriptscriptstyle S}}\xspace}

\def\Dbar    {\kern 0.2em\overline{\kern -0.2em \PD}{}\xspace}
\def\D       {\ensuremath{\PD}\xspace}

\def\Dz      {\texorpdfstring{\ensuremath{\D^0}}{D0}\xspace}
\def\Dzb     {\ensuremath{\Dbar^0}\xspace}
\def\DzDzb   {\ensuremath{\Dz {\kern -0.16em \Dzb}}\xspace}
\def\Dp      {\ensuremath{\D^+}\xspace}
\def\Dm      {\ensuremath{\D^-}\xspace}

\def\DpDm    {\ensuremath{\Dp {\kern -0.16em \Dm}}\xspace}

\def\B       {\ensuremath{\PB}\xspace}
  \def\Bbar    {\kern 0.18em\overline{\kern -0.18em \PB}{}\xspace}

\def\Bz      {\ensuremath{\B^0}\xspace}

\def\Bu      {\ensuremath{\B^+}\xspace}

\def\Bs      {\ensuremath{\B^0_\squark}\xspace}

  \def\Y#1S{\ensuremath{\PUpsilon{(#1S)}}\xspace}

\newcommand{\decay}[2]{\ensuremath{#1\!\to #2}\xspace}         

\def\to                 {\ensuremath{\rightarrow}\xspace}

\def\CP                {\ensuremath{C\!P}\xspace}

\def\AT#1     {\ensuremath{A_{\mathrm{T}}^{#1}}\xspace}           

\def\C#1      {\ensuremath{\mathcal{C}_{#1}}\xspace}                       
\def\Cp#1     {\ensuremath{\mathcal{C}_{#1}^{'}}\xspace}                    
\def\Ceff#1   {\ensuremath{\mathcal{C}_{#1}^{\mathrm{(eff)}}}\xspace}        
\def\Cpeff#1  {\ensuremath{\mathcal{C}_{#1}^{'\mathrm{(eff)}}}\xspace}       
\def\Ope#1    {\ensuremath{\mathcal{O}_{#1}}\xspace}                       
\def\Opep#1   {\ensuremath{\mathcal{O}_{#1}^{'}}\xspace}                    

\newcommand{\tev}{\ensuremath{\mathrm{\,Te\kern -0.1em V}}\xspace}
\newcommand{\gev}{\ensuremath{\mathrm{\,Ge\kern -0.1em V}}\xspace}
\newcommand{\mev}{\ensuremath{\mathrm{\,Me\kern -0.1em V}}\xspace}
\newcommand{\kev}{\ensuremath{\mathrm{\,ke\kern -0.1em V}}\xspace}
\newcommand{\ev}{\ensuremath{\mathrm{\,e\kern -0.1em V}}\xspace}
\newcommand{\gevc}{\ensuremath{{\mathrm{\,Ge\kern -0.1em V\!/}c}}\xspace}
\newcommand{\mevc}{\ensuremath{{\mathrm{\,Me\kern -0.1em V\!/}c}}\xspace}
\newcommand{\gevcc}{\ensuremath{{\mathrm{\,Ge\kern -0.1em V\!/}c^2}}\xspace}
\newcommand{\gevgevcccc}{\ensuremath{{\mathrm{\,Ge\kern -0.1em V^2\!/}c^4}}\xspace}
\newcommand{\mevcc}{\ensuremath{{\mathrm{\,Me\kern -0.1em V\!/}c^2}}\xspace}

\def\invfb   {\ensuremath{\mbox{\,fb}^{-1}}\xspace}

\def\gsim{{~\raise.15em\hbox{$>$}\kern-.85em
          \lower.35em\hbox{$\sim$}~}\xspace}
\def\lsim{{~\raise.15em\hbox{$<$}\kern-.85em
          \lower.35em\hbox{$\sim$}~}\xspace}

\def\tell1  {TELL1\xspace}
\def\ukl1   {UKL1\xspace}

\newcommand{\figs}{figs}

\newcommand{\omcl}{\ensuremath{1-{\rm CL}}\xspace}

\newcommand{\pdf}{\mbox{PDF}\xspace}

\newcommand{\robust}{robust\xspace}

\newcommand{\full}{full\xspace}

\renewcommand{\Kz}     {\texorpdfstring{\ensuremath{K^0}}{K0}\xspace}

\newcommand{\BDK}     {\texorpdfstring{\ensuremath{B\to DK}}			{B -> DK}\xspace}
\newcommand{\BDpi}    {\texorpdfstring{\ensuremath{B\to D\pi}}			{B -> Dpi}\xspace}

\newcommand{\BuDpi}   {\texorpdfstring{\ensuremath{\Bu\to D \pip}}		{B+ -> Dpi+}\xspace}
\newcommand{\BuDK}    {\texorpdfstring{\ensuremath{\Bu\to D \Kp}}		{B+ -> DK+}\xspace}
\newcommand{\BuDh}    {\texorpdfstring{\ensuremath{\Bu\to D h^+}}		{B+ -> Dh+}\xspace}

\newcommand{\BdDKstz} {\texorpdfstring{\decay{\Bz}{D K^{*0}}}		{}}

\newcommand{\BsDsK}	{\texorpdfstring{\ensuremath{\Bs\to D_s^\mp K^\pm}}	{Bs -> DsK}\xspace}

\newcommand{\DKpipipi} {\texorpdfstring{\ensuremath{D\to K\pi\pi\pi}}{D -> K3pi}\xspace}

\newcommand{\DKSKpi}   {\texorpdfstring{\ensuremath{D\to \KS K\pi}}{D -> KSKpi}\xspace}

\newcommand{\Dhh}      {\texorpdfstring{\ensuremath{D\to hh}}{D -> hh}\xspace}

\newcommand{\DKShh}    {\texorpdfstring{\ensuremath{D\to\KS hh}}{D -> KShh}\xspace}

\renewcommand{\g}{\texorpdfstring{\ensuremath{\gamma}}{gamma}\xspace}

\newcommand{\rbpi}  {\texorpdfstring{\ensuremath{r_B^{D\pi}}}{rBDpi}\xspace}

\newcommand{\gQuoted}                                   {\ensuremath{73^{+9}_{-10}}\xspace}

\newcommand{\gRobustCentral}    {\ensuremath{72.9}\xspace}

\newcommand{\gRobustOnesig}     {\ensuremath{[63.0,82.1]}\xspace}
\newcommand{\gRobustTwosig}     {\ensuremath{[52.0,90.5]}\xspace}

\newcommand{\gRobustCentralPMProb}  {\ensuremath{\gRobustCentralPMProb^{+8.3}_{-9.2}}\xspace}

\newcommand{\gFullCentralA}     {\ensuremath{78.9}\xspace}
\newcommand{\gFullCentralB}     {\ensuremath{72.8}\xspace}

\newcommand{\gFullOnesig}       {\ensuremath{[71.5,84.7]}\xspace}
\newcommand{\gFullTwosig}       {\ensuremath{[54.6,91.4]}\xspace}

\newcommand{\rbpiFullCentralA}  {\ensuremath{0.027}\xspace}
\newcommand{\rbpiFullCentralB}  {\ensuremath{0.006}\xspace} 
\newcommand{\rbpiFullOnesigA}   {\ensuremath{[0.016,0.034]}\xspace}
\newcommand{\rbpiFullOnesigB}   {\ensuremath{[0.005,0.007]}\xspace} 
\newcommand{\rbpiFullTwosig}    {\ensuremath{[0.001,0.040]}\xspace}

\newcommand{\gRobustCentralBayes}    {\ensuremath{71.9}\xspace}

\newcommand{\gRobustOnesigBayes}     {\ensuremath{[61.9,81.8]}\xspace}
\newcommand{\gRobustTwosigBayes}     {\ensuremath{[50.9,91]}\xspace}

\begin{document}
\begin{titlepage}
\pubblock

\vfill
\Title{Improved constraints on \g: \\ CKM2014 update}
\vfill
\Author{Till Moritz Karbach\support}
\Address{\napoli}
\vfill
I report on an updated combination of all currently available tree-level measurements of
the CKM angle $\gamma$ from \lhcb. This combination currently gives LHCb's most precise
value of $\g = (\gQuoted)^\circ$ obtained from \BDK-like decays. For the first time in a single experiment,
the uncertainty has dropped below $10^\circ$. When using \BDpi decays in addition, the
situation becomes non-Gaussian which makes the confidence intervals more difficult to interpret.
\vfill
\begin{Presented}
8th International Workshop on the CKM Unitarity Triangle (CKM 2014),
Vienna, Austria, September 8-12, 2014
\end{Presented}
\vfill
\end{titlepage}
\def\thefootnote{\fnsymbol{footnote}}
\setcounter{footnote}{0}

\section{Introduction}

The CKM angle $\gamma = \arg \left[ -V_{ud}V_{ub}^*/(V_{cd}V_{cb}^*) \right]$
is a central parameter describing the CKM matrix. It can be measured using two
different approaches. One is using loop-induced decays, which can give rise to 
effects of New Physics. The other approach is the topic of these proceedings
and is purely based on tree level decays.
This last property turns \g into one of the corner stones of CKM physics.

At LHCb, \g is measured using a large variety of decay channels. The best sensitivity
is achieved through a combination of measurements that determine \g along with
several other hadronic parameters. The input measurements provide sensitivity to \g through the interference
of $b\to u$ and $b\to c$ amplitudes, as described in more detail in Ref.~\cite{Aaij:2013zfa}. 
Determining the additional hadronic parameters from data results
in a small systematic uncertainty on the measurement of \g.
The presented update~\cite{LHCb-CONF-2014-004} improves over the previous
combinations~\cite{Aaij:2013zfa,LHCb-CONF-2013-006} by
adding more decay channels and updating selected channels to the full
available dataset of 3\invfb. Tab.~\ref{tab:inputs} gives the input measurements 
considered in this combination. Among the inputs is the first constraint
of \g from the time-dependent analysis of \BsDsK~\cite{Aaij:2014fba}.
Being the only time-dependent method considered so far at LHCb, it is affected by
very different experimental systematic uncertainties, which bases the \g average
on a healthy range of measurements.

\begin{table}[!h]
\centering
\caption{LHCb Input measurements. The symbol $h$ denotes either a kaon or a pion,
and the acronyms denote the initials of authors first proposing
the measurements in Refs.~\cite{Gronau:1991dp,Gronau:1990ra,Atwood:1996ci,Atwood:2000ck,
Bondar:2001vn,Giri:2003ty,Grossman:2002aq,Fleischer:2003yb,Aleksan:1991nh,Dunietz:1987bv}.}
\label{tab:inputs}
\begin{tabular}{lccc}
\hline
\BuDh, \Dhh & GLW/ADS & 1\invfb & \cite{Aaij:2012kz}\\
\BuDh, \DKpipipi& ADS& 1\invfb& \cite{Aaij:2013mba}\\
\BuDK, \DKShh& model-independent GGSZ& 3\invfb&\cite{Aaij:2014iba}\\
\BuDK, \DKSKpi& GLS& 3\invfb&\cite{Aaij:2014dia}\\
\BdDKstz, \Dhh& GLW/ADS& 3\invfb&\cite{Aaij:2014eha}\\
\BsDsK& time-dependent& 1\invfb&\cite{Aaij:2014fba}\\
\hline
\end{tabular}
\end{table}

\section{Two combinations}

Two combinations are prepared, referred to as the ``\robust'' and ``\full''
combinations. The robust combination only contains observables measured
in \BDK-like systems, which are the traditional channels being used to measure
\g. Compared to \BDpi-like systems they offer larger interference, and therefore are very robust
against systematic effects. The robust combination provides the main result
for \g.

The \full combination adds
information from the \BDpi system.
This is motivated by the fact that LHCb has a large set
of \g-sensitive observables in \BDpi decays, although their
sensitivity to \g is suppressed compared to the \BDK-like decays. The reason for this is that
the amplitude ratio, which governs the interference effects and therefore
the sensitivity, is expected to be a factor $\approx 15$ smaller than for
the \BDK-like systems.
At the same time, the available data samples are usually a factor
10 larger in \BDpi decays. The \full combination
is more sensitive to several effects. The most prominent of these is \Dz--\Dzb mixing~\cite{Meca:1998ee,Silva:1999bd,Rama:2013voa},
which we fully correct for in both combinations, taking into account the
\Dz decay time acceptances of the individual measurements.
We also correct at first order for possible \CP violation in the \Dz system
using information from the Heavy Flavor Averaging Group
(HFAG)~\cite{HFAG2012}. Furthermore, in order to ensure our combination is not sensitive to mixing
and \CP violation in \Kz decays~\cite{Grossman:2013woa}, we exclude observables where such effects may be non-negligible.
Due to the
larger data samples, the relative impact of the systematic uncertainties is larger than in the robust combination,
although both of the current combinations are still statistically limited.
Finally, a small value of the amplitude ratio \rbpi that is not significantly different from
zero, is known to affect the
coverage of the frequentist methods used here.

\section{Statistical Procedure}

The combination follows a frequentist treatment described in detail in Ref.~\cite{Aaij:2013zfa},
in which the nuisance parameters
are kept at their best fit values (known as the ``plug-in'' method).
The results obtained using \BDK-like decays
alone are also cross-checked using a Bayesian approach assuming a flat prior.

\section{Results}

\begin{figure}[htb]
  \centering
  \includegraphics[width=.45\textwidth]{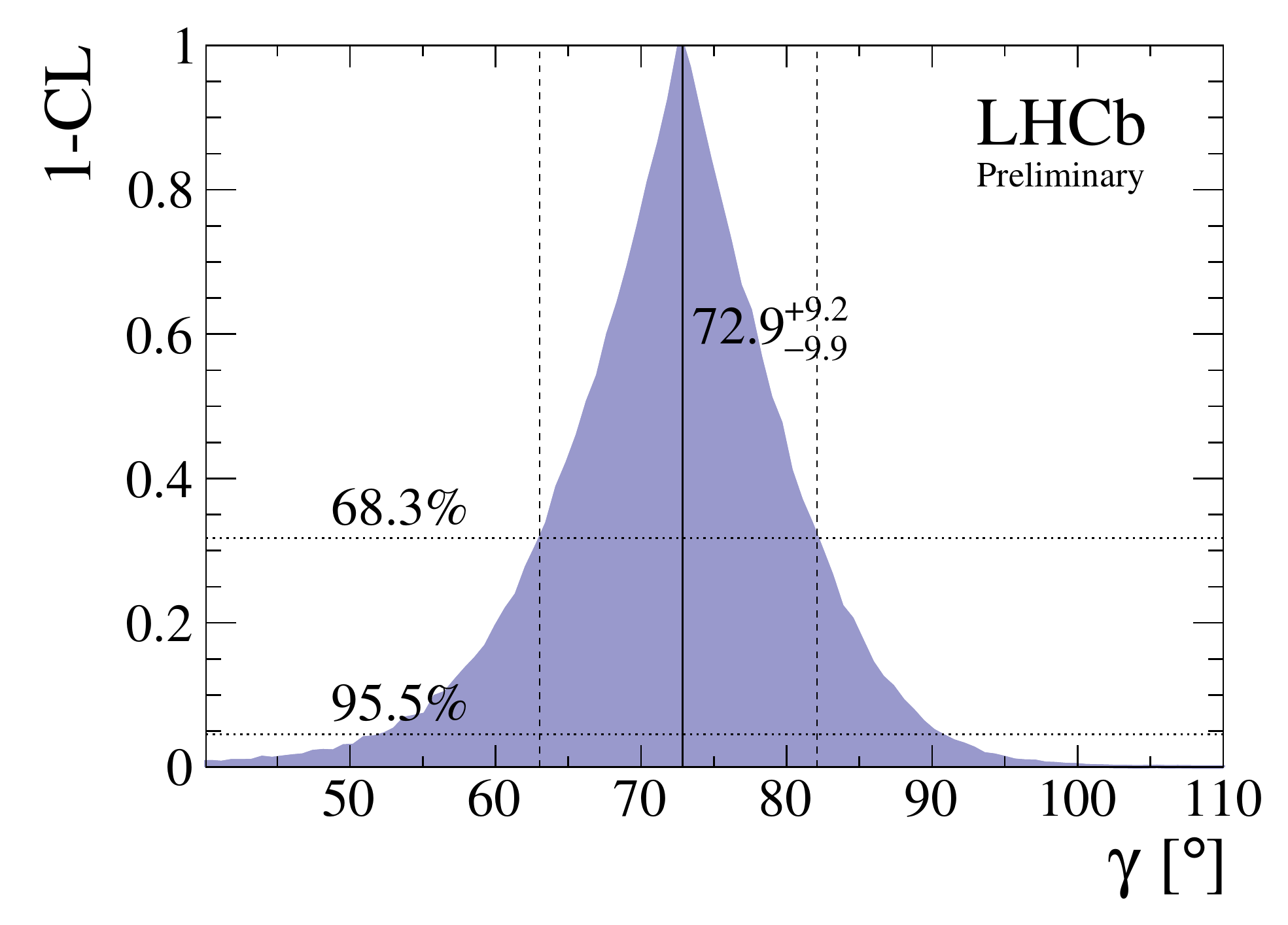}
  \includegraphics[width=.45\textwidth]{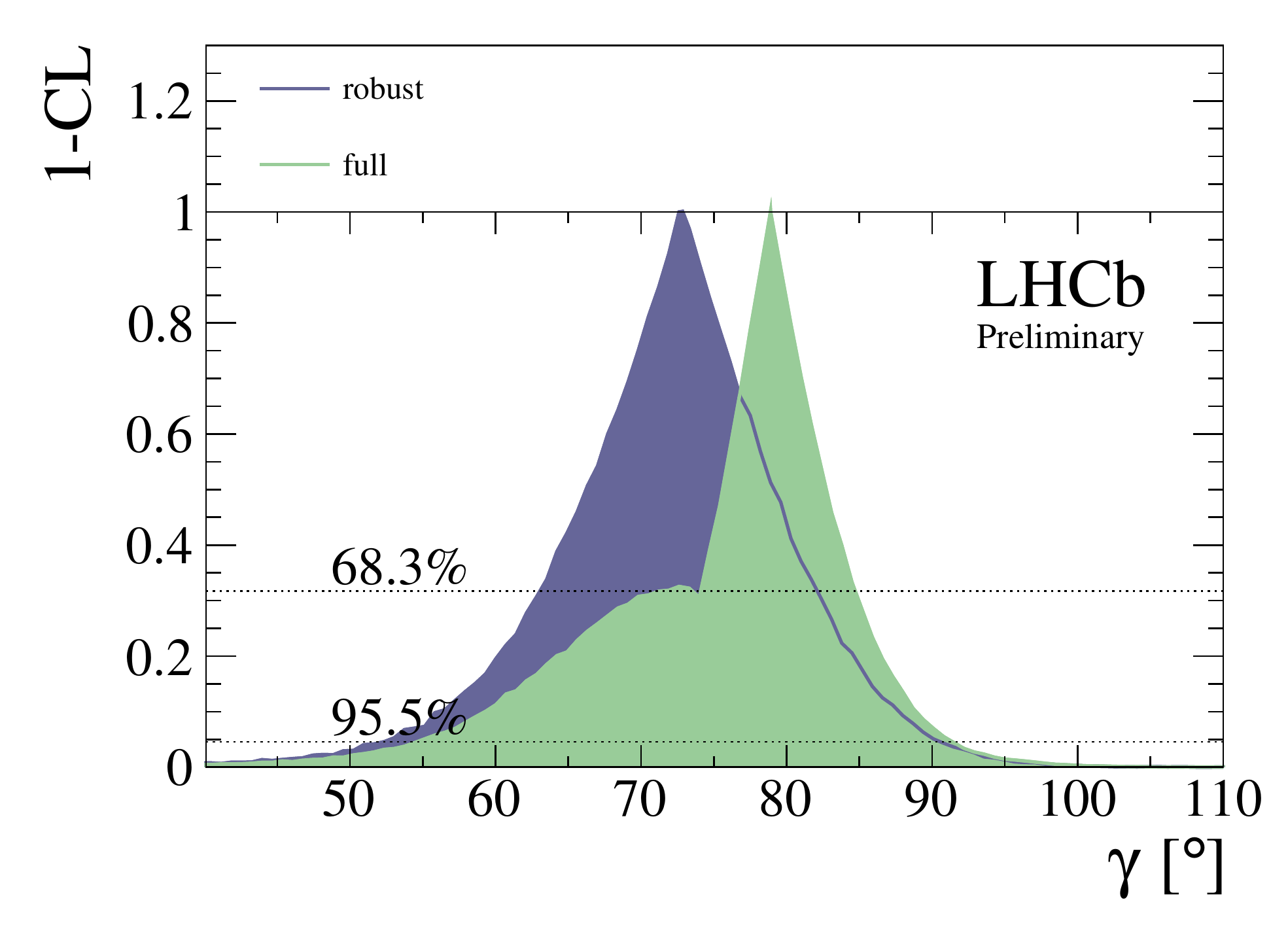}
  \caption{\omcl curves for the \robust combination (left), and for
  both \robust and \full combinations (right).
  }
  \label{fig:resultfreq}
\end{figure}

\begin{figure}[h!]
\begin{center}
\includegraphics[trim={0 0 0 12cm},clip,width=0.49\textwidth]{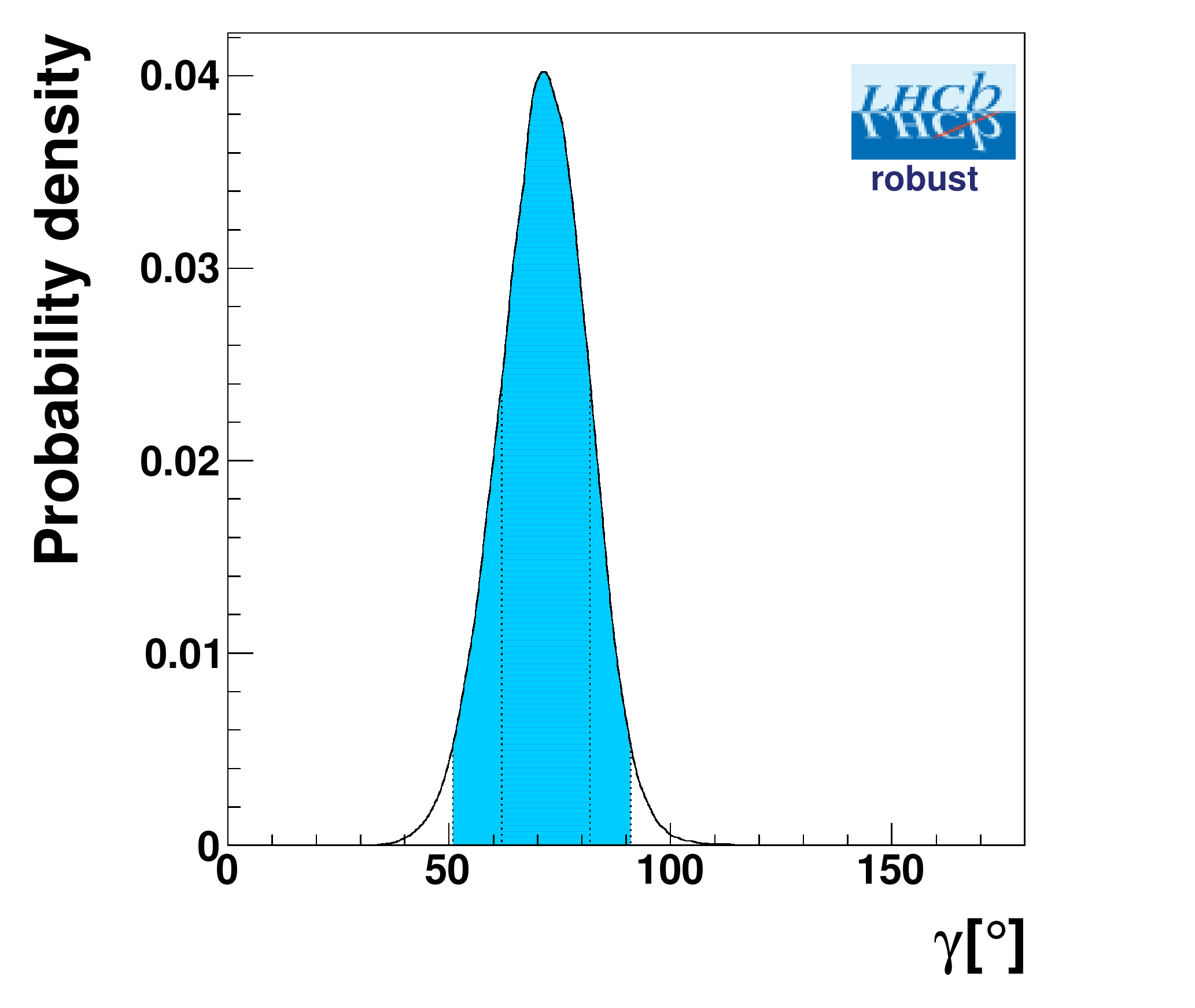}
\end{center}
\caption{The 1D posterior \pdf\ in the \robust combination. Dark and light regions show the 68\% 
and 95\% probability intervals, respectively.
\label{fig:DKcombBayes}}
\end{figure}

The result of the robust combination is given in 
Tab.~\ref{tab:resultrobust}, and illustrated in Figs~\ref{fig:resultfreq}
and Fig~\ref{fig:DKcombBayes}. Both frequentist and Bayesian 
intervals agree very well. The coverage was tested for the frequentist interval and found
to be accurate. The fit probability
is $89.4\%$. The robust frequentist values are regarded as the
nominal result: $\g = (\gQuoted)^\circ$. 

\begin{table}[!htb]
\caption{\label{tab:resultrobust}Confidence and credibility regions 
and central values for \g extracted from the \robust combination.
}
\begin{center}
\begin{tabular}{lcccc}
\hline
         & Observable & Central value & \multicolumn{2}{c}{Intervals}\\
         &            &               & 68\%  &95\%  \\
\hline
Frequentist & $\gamma [^{\circ}]$     & \gRobustCentral       & \gRobustOnesig        & \gRobustTwosig \\
Bayesian    & $\gamma [^{\circ}]$      & \gRobustCentralBayes & \gRobustOnesigBayes   & \gRobustTwosigBayes \\
\hline
\end{tabular}
\end{center}
\end{table}

The full combination is more difficult to interpret.
It exhibits a sharp maximum at $\g=\gFullCentralA^\circ$,
and a secondary maximum at a value similar to the maximum of the \robust
combination (Tab.~\ref{tab:resultfull}). This sharp maximum corresponds to an unexpectedly large value
of the amplitude ratio \rbpi, $\rbpi=\rbpiFullCentralA$. This is regarded as a fluctuation,
but nevertheless enhances the impact of the \BuDpi system. It makes the
situation visibly non-Gaussian. As a consequence,
the 68\% CL interval is misleadingly small, while at 
95\% CL the intervals agree well between both combinations.
This is illustrated in Fig.~\ref{fig:resultfreq} (right), which
shows both \robust and \full combinations in the same plot.
The coverage of the intervals of the \full combination was also tested,
and as expected from the fact that \rbpi is still consistent with zero,
the intervals were found to undercover a bit, i.e.~the reported intervals
are too small. It is expected that the ambiguity in \rbpi, which causes
the non-Gaussian behavior, could be resolved by a GGSZ-type analysis of
\BuDpi decays.

\begin{table}[!h]
\centering
\caption{Confidence intervals and central values
for the \full combination. The two columns correspond to the two minima found by the fit. The
most probable value is given in the left column, corresponding to a large value of \rbpi.}
\label{tab:resultfull}
\begin{tabular}{lcc}
\hline
quantity & \multicolumn{2}{c}{\full}                            \\
\hline
\g ($^\circ$)      & \gFullCentralA & \gFullCentralB    \\
68\% CL ($^\circ$) & \multicolumn{2}{c}{\gFullOnesig}   \\
95\% CL ($^\circ$) & \multicolumn{2}{c}{\gFullTwosig}   \\
\hline
\rbpi           & \rbpiFullCentralA & \rbpiFullCentralB                 \\
68\% CL             & \rbpiFullOnesigA  & \rbpiFullOnesigB      \\
95\% CL                 & \multicolumn{2}{c}{\rbpiFullTwosig}                    \\
\hline
\end{tabular}
\end{table}

\section{Discussion}

Following the presentation there was a discussion about the following points:
\begin{itemize}
\item Why is the \BDK-like ``robust'' combination regarded as the nominal
result, rather than the more complete ``full'' combination? --- At this point,
LHCb doesn't have enough data to measure \rbpi significantly different from
zero, which affects the statistical coverage of the ``full'' combination.
The result is still correct, only much more difficult to interpret. But the \BuDpi
system is a perfect test bed to show up the obstacles on the way to a
high precision measurement of \g in \BuDK.

\item What drives the large \rbpi minimum? --- It is driven by two analyses
taken together: \BuDh, \Dhh~\cite{Aaij:2012kz} and \BuDh, \DKpipipi~\cite{Aaij:2013mba}.
If either one is dropped, the fit converges to a lower value of \rbpi. 

\item Does LHCb plan to make a GGSZ type measurement in \BuDpi? --- Yes.

\item Why did the precision on \g decrease by so much when the auxiliary 
inputs were updated? --- This was tracked down to a sizable shift in the
parameter $R_D(K\pi)$ in the latest HFAG average compared to previous ones.

\end{itemize}

\section{Conclusion}

Observables measured by LHCb that have sensitivity at tree-level to the
CKM angle \g, along with supplementary information from other experiments,
are combined to determine an improved constraint on \g.
The effect of \Dz--\Dzb mixing on the decay rate
is taken into account, with consideration of the experimental decay time acceptances
of the individual measurements.
When all observables are combined, we find
\begin{align}
\g = (\gQuoted)^\circ\,,
\end{align}
using \BDK-like decays only.
This result is more precise than the combination of the results of the $B$ factories
Babar and Belle~\cite{Bevan:2014iga}.

\addcontentsline{toc}{section}{References}
\bibliographystyle{LHCb}
\bibliography{references,bayesian}

\end{document}